# Temperature Dependent Local Structure of $LaFeAsO_{1-x}F_x$: Probing the Bond Correlations

T. A. Tyson[1,2], T. Wu[1], J. Woicik[3], B. Ravel[3], A. Ignatov[2], C. L. Zhang[2], Z. Qin[3], T. Zhou[1,2] and S.-W. Cheong[2]

[1]Department of Physics, New Jersey Institute of Technology, Newark, NJ 07102
[2]Department of Physics and Astronomy & Rutgers Center for Emergent Materials, Rutgers University, Piscataway, NJ 08854 and
[3]National Institute of Standards and Technology, Gaithersburg, Md 20899

**ABSTRACT**

The local structure of the parent and doped $LaFeAsO_{1-x}F_x$ (pnictide) compounds were studied by x-ray absorption spectroscopy. In the doped system, the Fe-As and Fe-Fe correlations are well modeled by an Einstein model with no low temperature anomalies. For the Fe-As bonds, the Einstein temperatures are identical for the doped (11%) and undoped samples, but the doped sample is found to have a lower level of static disorder. For the Fe-Fe correlation, doping enhances the effective Einstein temperature of Fe-Fe atom correlation. The results suggest that the onset of superconductivity in the F doped system may be related to enhanced magnetic correlations. Density functional calculations of the charge density reveal strong bonding between neighboring As ions but metal-like behavior in the Fe layers.

PACS: 74.70-b, 78.70.Dm, 61.05.cj

With the discovery of superconductivity in the non copper based pnictide system $LaFeAsO_{1-x}F_x$ [1, 2], much work is being conducted to unravel the mechanism behind superconductivity in this material. A fundamental understanding of the structure from the perspective of the long range (unit cell averaged) and local (local site averaged) is essential to providing the details needed to develop a microscopic level model. The parent compound (x=0) in $LaO_{1-x}F_xFeAs$ exhibits a transition from an orthorhombic phase (P4/nmm) to a tetragonal phase (Cmma) at low temperature near ~160 K followed by antiferromagnetic (AFM) ordering at ~145 K [3]. The most recent measurements indicate that the onset of the orthorhombic distortion occurs at ~200 K with an abrupt increase near ~160 K [2(c)]. In the underdoped system, orthorhombic distortions is found to exist beyond the AFM phase and coexists with the superconducting phase with no evidence for static long range AFM below the superconducting temperature [3(a)]. Doping with F suppresses the splitting of the (400) and (040) beyond x=0.05. The result suggest competition between AFM order and superconductivity. Diffraction based measurements [3] indicate that F doping compresses c and makes a approach b (with respect to the orthorhombic cell). Doping with F brings the ReO and FeAs block closer. No abrupt changes in structure in the low temperature phase is seen with doping. The resulting structural changes with doping make the Fe-As-Fe bond angle approach the value appropriate for a perfect tetrahedron of 109.5°. Local structural measurement may provide insight which will complement the unit cell averages structure obtained by neutron and x-ray diffraction.

To directly probe changes in the local structure about the Fe sites x-ray absorption measurements were conducted between 15 and 300 K in the parent and doped (11%) $LaFeAsO_{1-x}F_x$. In the doped system, the Fe-As and Fe-Fe correlations are well modeled by an Einstein model with no low temperature anomalies. While the Einstein temperatures are identical for the doped (11%) and undoped samples, the doped sample is found to have a lower level of static disorder in the Fe-As distribution. For the Fe-Fe correlation, doping enhances the effective Einstein temperature. The results suggest that the onset of superconductivity in the F doped system may be related to enhanced magnetic correlations.

Polycrystalline samples of were prepared by solid state reaction as described in Ref. [4]. Rietveld refinement of x-ray diffraction data on the $x = 0$ system reveal no detectable impurities while LaOF impurity levels at ~4% were detected in the x=0.11 sample. We note also that in the in F doped system the exact F doping is difficult to determine. However, in our doped sample the resisistivity data indicate an onset of superconductivity near 23 K.

X-ray absorption samples were prepared by grinding and sieving the material (500 mesh) and brushing it onto Kapton tape. Layers of tape were stacked to produce a uniform sample for transmission mode measurements. Spectra were measured at the NSLS beamline X23A2 at Brookhaven National Laboratory. Measurements were made on warming from 15 K on a closed-cycle Displex™ cryostat. Two to four scans were taken at each temperature. The uncertainty in temperature is < 0.1 K. A Fe foil reference was employed for energy calibration. The reduction of the x-ray absorption fine-structure (XAFS) data was performed using standard procedures [5]. Representative XAFS data at 160 K are shown in Fig. 1(a)

To treat the atomic distribution functions on equal footing at all temperatures the spectra were modeled in R-space by optimizing the integral of the product of the radial distribution functions, representing each unique shell, and theoretical spectra with respect to the measured spectra (as in Ref. [6]). Theoretical spectra for atomic shells [7] were derived from the P4/nmm crystal structure [8]. Both the Fe-As and Fe-Fe (nearest neighbor) distributions were modeled: k-range 2.2 < k < 16.2 Å$^{-1}$ and r-range 1.54 < R < 3.00 Å . The coordination numbers of the two shells were held at N1=N2=4 and Gaussian widths ($\sigma^2 = \langle (R - \langle R \rangle)^2 \rangle$ giving the mean squared relative displacement of a bond)) and bond distances for the Fe-As and Fe-Fe shells are determined. Typical XAFS data (x=0) are shown as multiple scans at 165 K in Fig.1(a) and a typical two shell- fit in R-space is shown in Fig. 1(b). The extracted width were fit to a simple Einstein model $\sigma^2(T) = \sigma_0^2 + \frac{\hbar^2}{2\mu k_B \theta_E} \coth(\frac{\theta_E}{2T})$ [9], where μ is the reduced mass for the bond pair and a parameter $\sigma_0^2$ represents the static disorder. This simple model represents the bond vibrations as harmonic oscillations of a single effective frequency proportional to $\theta_E$. It provides an approach to characterize the relative stiffness of the bonds. The static parameter $\sigma_0^2$ for the Fe-Fe correlation was found to be approximately zero. Errors bars shown represent the statistical uncertainty in the data.

To understand the difference in correlation functions for the Fe and As ions the charge density as examined. Calculation of the charge density within the local spin density approximation (U/J=0) was carried out using highly accurate all-electron full potential linear augmented plane-wave plus local orbital (FPLAPW+lo) method implemented in the WIEN2k code [10]. The parent compound structure at 2 K from Ref. [3(a)] was used.

In Fig. 2 we show the temperature dependence of the Fe-As correlation as a function of temperature for the doped and the undoped system. The Fe-As bond correlation is fit well for both the doped and undoped samples over the entire temperature range with a simple two parameter ($\sigma_0^2$ and $\theta_E$) model of the vibration of the atom pair. While the effective Einstein temperature for the bond is the same for both the doped and undoped system, there in a decrease in the $\sigma_0^2$ static disorder of the bond with F doping.

The Fe-Fe correlations are best modeled the $\sigma_0^2 = 0$ indicating a very low degree of static disorder with respect to this bond pair (see Fig. 3). While the 11% doped system is fit well over the entire temperature range by an Einstein model, significant variation from this model are found at low temperature for the parent compound. Specifically below ~100K, there is a reduction of $\sigma^2$ for the undoped system. In fact the points fall on the curve for the doped system. Note also that in the region of the structural phase transition (~160K) no significant changes are seen showing the there is no large local distortion in the x=0 system in transiting from the tetragonal to orthorhombic phase. The Einstein temperature for the doped system is significantly higher than that for the undoped system (304(2) K compared to 269(2) K).

To qualitatively understand the difference between in the Fe-Fe and Fe-As correlations we show the charge density (with respect to the orthorhombic unit cell [3(a)]) for the parent compound as constant z plots in Fig. 4(a) and 4(b) yielding charge density in the As and Fe layers, respectively.. In Figs. 4(c) and 4(d) we plot the charge density for y=0 and y=0.25 planes. Note while the As ions sites are covalently bonded to the Fe sites and also to the neighboring As sites (As-As hybridization) with highly anisotropic charge along bonding directions, the planes containing the Fe ions reveal metallic like-charge distribution with low asymmetry (Figs. 4(b) and 4(c)). Apart from the effect of the small difference in distance between the Fe-As and Fe-Fe bond, the reduced Einstein temperature is due to the weaker bonding in the Fe plane. On a local level this bonding is enhanced at low temperatures suggesting that strong short-range magnetic correlations exist in this system even in the doped superconducting material.

Put together we make the following observations. At the level of the experiments, the doped system is well modeled by Einstein type pair correlations for the Fe-As and Fe-Fe bonds. For the undoped system the Fe-Fe bond are also simply modeled but possibly with two distinct regions T < 100 K and T > 100 K. From a local perspective, the primary impact of doping on a local atomic scale is an enhancement of the Fe-Fe correlation and reduction of static disorder. The results are consistent with magnetic origin of superconductivity proposed in theories of these materials (see for example Refs [11, 12, 13, 14]).

This work is supported by DOE Grants DE-FG02-07ER46402 (NJIT) and DE-FG02-07ER46382 (Rutgers University). Data acquisition was performed at Brookhaven National Laboratory's National Synchrotron Light Source (NSLS) which is funded by the U. S. Department of Energy.

# Figure Captions

**(Color online) Fig. 1**. (a) Four consecutive scans at 165 K (x=0) showing the quality of the Fe K-Edge XAFS data. (b) Fourier transform of XAFS data taken at 165K K. The dashed thin lines are fits to the data. The first and second shell Fe-As and Fe-Fe peaks are labeled. Peaks are shifted to lower R-values from real distances by atomic phase shifts.

**(Color online) Fig. 2**. Extracted $\sigma^2(T)$ for the Fe-As bond for the x=0 amd x=0.11 samples as the upper and lower curve. Note the Einstein ($\theta_E$) temperature is the same for both samples.

**(Color online) Fig. 3**. Extracted $\sigma^2(T)$ for the Fe-Fe bonds for the x=0 amd x=0.11 sample as the upper and lower curve. Note the Einstein ($\theta_E$) temperature for this correlation is higher for the doped sample indicating enhanced Fe-Fe coupling with F doping.

**(Color online) Fig. 4**. In panels (a) and (b) the change density in the planes for z=0.65 and z=0.5 are given showing the charge distribution in the As and Fe layers, respectively. The planes at y=0 and y=0.25 are shown in (c) and (d). All plots have the same z scale (as in (a)) in units of electrons/Bohr radius cubed..

**Fig. 1. Tyson *et al*.**

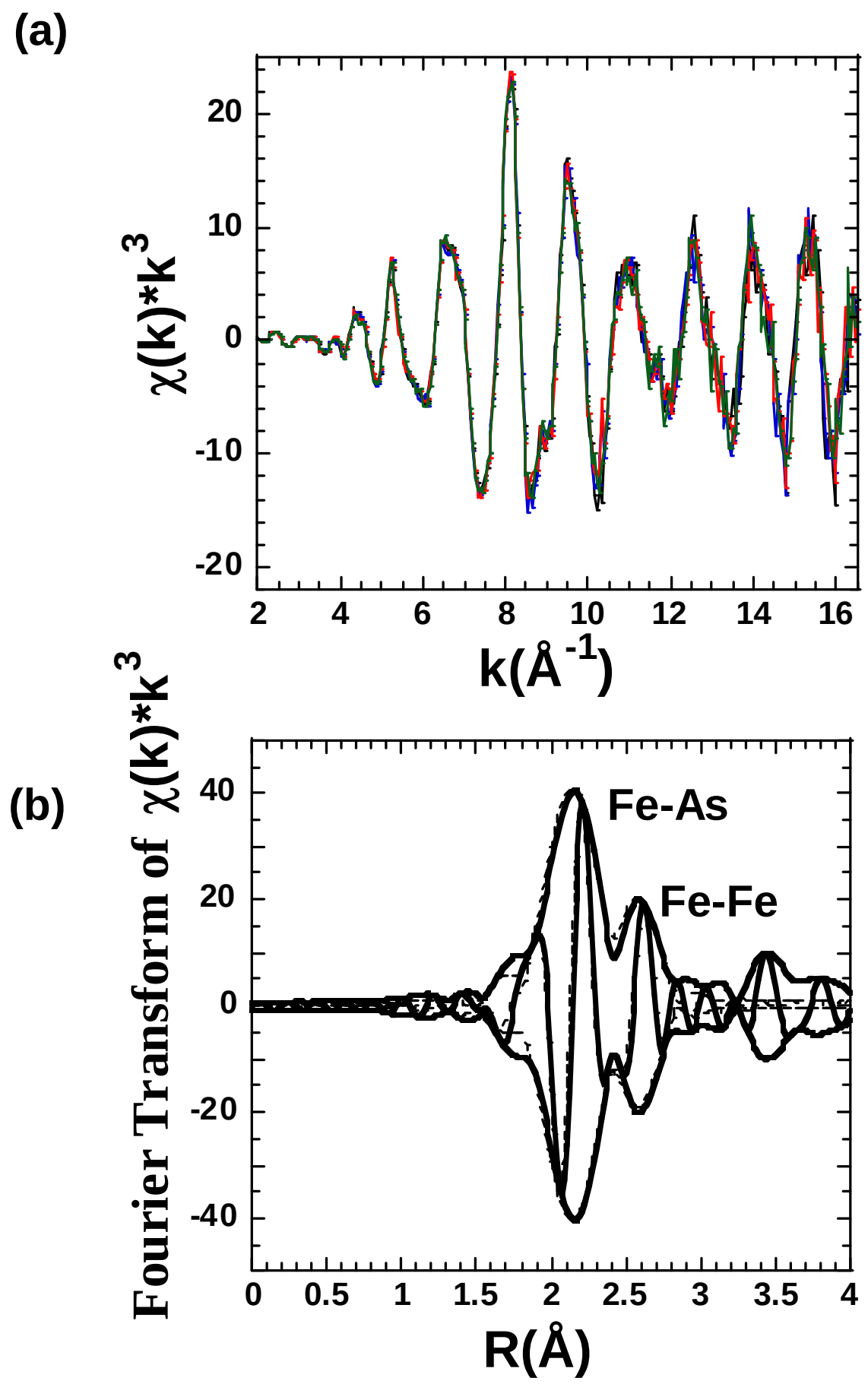

**Fig. 2. Tyson *et al*.**

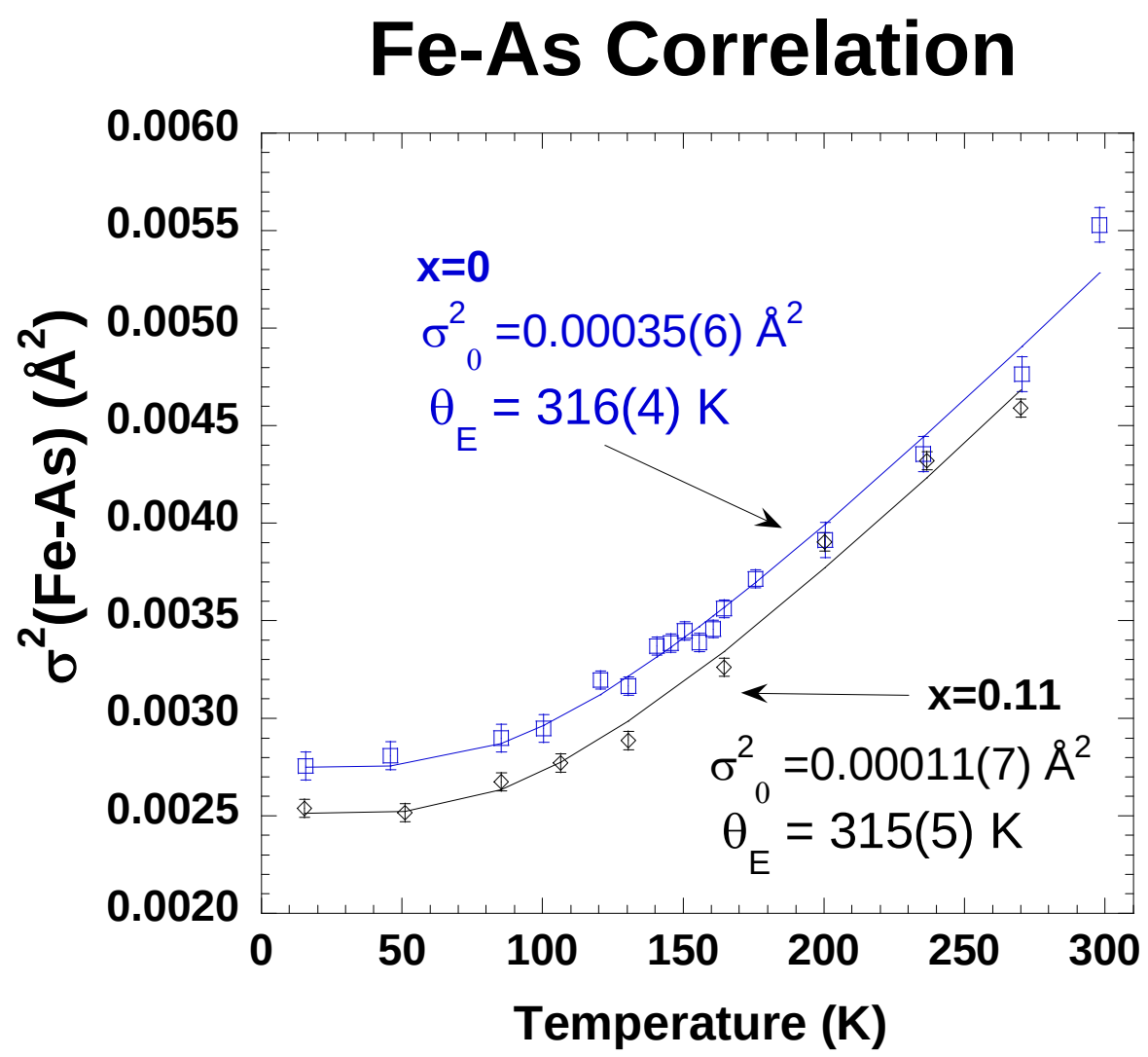

**Fig. 3. Tyson *et al*.**

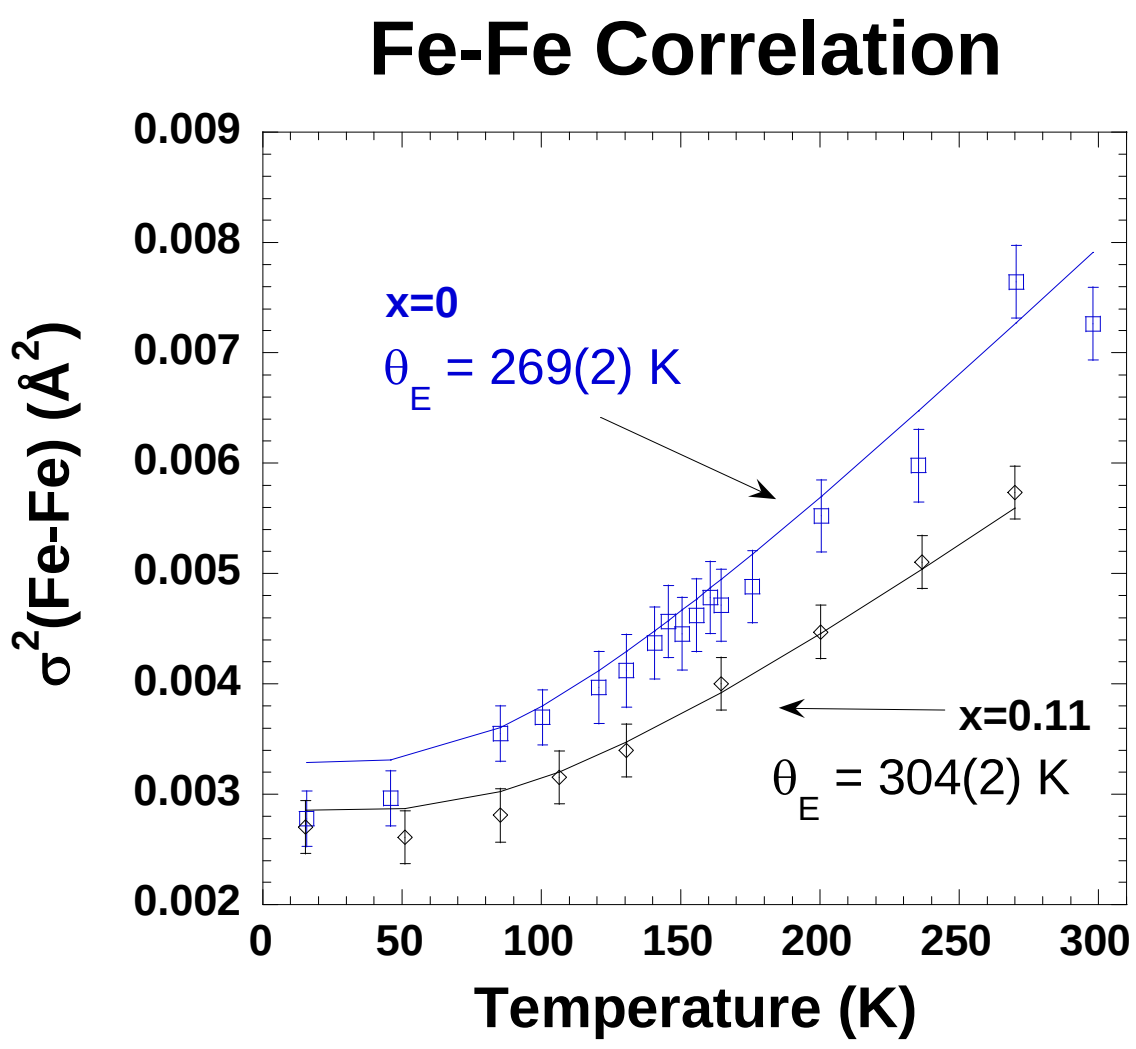

**Fig. 4. Tyson *et al***

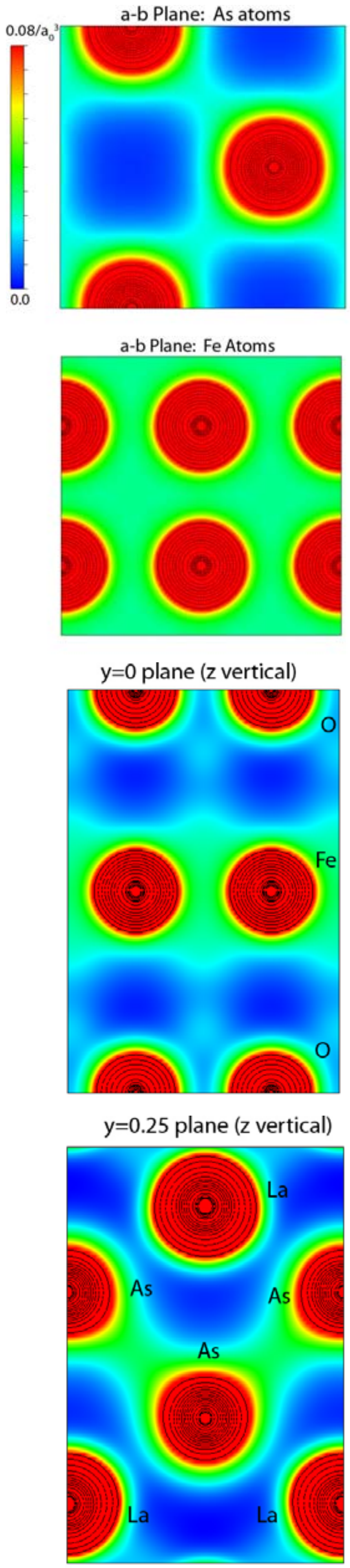

**(a)**

**(b)**

**(c)**

**(d)**